\documentclass{article}

\usepackage{arxiv}
\usepackage{graphicx}
\usepackage{tabularx}
\usepackage{multirow}
\usepackage{multicol}
\usepackage{array,xcolor,colortbl}
\usepackage{url}

\usepackage[utf8]{inputenc} % allow utf-8 input
\usepackage[T1]{fontenc}    % use 8-bit T1 fonts
\usepackage{hyperref}       % hyperlinks
\usepackage{url}            % simple URL typesetting
\usepackage{booktabs}       % professional-quality tables
\usepackage{amsfonts}       % blackboard math symbols
\usepackage{nicefrac}       % compact symbols for 1/2, etc.
\usepackage{microtype}      % microtypography
\usepackage{lipsum}
\pdfoutput=1

\title{Chest x-ray automated triage: a semiologic approach designed for clinical implementation, exploiting different types of labels through a combination of four Deep Learning architectures.}

\author{
Candelaria Mosquera,\textsuperscript{1,2,*}
Facundo Diaz,\textsuperscript{3}
Fernando Binder,\textsuperscript{1}
José Martín Rabellino,\textsuperscript{3}\\
\bf Sonia Elizabeth Benitez,\textsuperscript{1}
\bf Alejandro Beresñak,\textsuperscript{3}
\bf Alberto Seehaus,\textsuperscript{3}
\bf Gabriel Ducrey,\textsuperscript{3}\\
\bf Jorge Alberto Ocantos,\textsuperscript{3}
\bf Daniel Roberto Luna,\textsuperscript{1}\\\\
\textsuperscript{1}{Health Informatics Department, Hospital Italiano de Buenos Aires, Argentina}\\
\textsuperscript{2}{Universidad Tecnológica Nacional, Argentina}\\
\textsuperscript{3}{Radiology Department, Hospital Italiano de Buenos Aires, Argentina}
}
\begin{document}
\maketitle
\begin{abstract}
\textbf{Background and Objectives:} The multiple chest x-ray datasets released in the last years have ground-truth labels intended for different computer vision tasks, suggesting that performance in automated chest-xray interpretation might improve by using a method that can exploit diverse types of annotations. This work presents a Deep Learning method based on the late fusion of different convolutional architectures, that allows training with heterogeneous data with a simple implementation, and evaluates its performance on independent test data. We focused on obtaining a clinically useful tool that could be successfully integrated into a hospital workflow.\\
\textbf{Materials and Methods:} Based on expert opinion, we selected four target chest x-ray findings, namely lung opacities, fractures, pneumothorax and pleural effusion. For each finding we defined the most adequate type of ground-truth label, and built four training datasets combining images from public chest x-ray datasets and our institutional archive. We trained four different Deep Learning architectures and combined their outputs with a late fusion strategy, obtaining a unified tool. Performance was measured on two test datasets: an external openly-available dataset, and a retrospective institutional dataset, to estimate performance on local population. \\
\textbf{Results:} The external and local test sets had 4376 and 1064 images, respectively, for which the model showed an area under the Receiver Operating Characteristics curve of 0.75 (95$\%$CI: 0.74-0.76) and 0.88 (95$\%$CI: 0.86-0.89) in the detection of abnormal chest x-rays. For the local population, a sensitivity of 86$\%$ (95$\%$CI: 84-90), and a specificity of 88$\%$ (95$\%$CI: 86-90) were obtained, with no significant differences between demographic subgroups. We present examples of heatmaps to show the accomplished level of interpretability, examining true and false positives. \\
\textbf{Conclusion:} This study presents a new approach for exploiting heterogeneous labels from different chest x-ray datasets, by choosing Deep Learning architectures according to the radiological characteristics of each pathological finding. We estimated the tool’s performance on local population, obtaining results comparable to state-of-the-art metrics. We believe this approach is closer to the actual reading process of chest x-rays by professionals, and therefore more likely to be successful in a real clinical setting. 
\end{abstract}

% keywords can be removed
\keywords{radiography \and artificial intelligence \and deep learning \and clinical decision support systems \and chest}

\textsuperscript{*}\textit{Correspondence author: candelaria.mosquera@hospitalitaliano.org.ar. Juan Domingo Perón 4190,}\\ \textit{C1199AAB, Ciudad Autónoma de Buenos Aires, Argentina. Tel (+54) 01149590200 - Int 505}
\section{Introduction}
The chest radiography (CXR) is one of the most commonly performed and well-established imaging modalities, playing an important role in diagnosis and monitoring of primary care \cite{Folio2012-lc}. At the same time, the increasing clinical demand on radiology departments worldwide is challenging current service delivery models, particularly in publicly funded health care systems. In some settings, it may not be feasible to report all acquired radiographs promptly, leading to large backlogs of unreported studies \cite{Clinical_Excellence_Commission2014-uc,Cliffe2016-fm}. Alternative models of care should be explored, particularly for chest radiographs, which account for 40\% of all diagnostic images worldwide \cite{World_Health_Organization2016-dt}. Automated CXR interpretation could improve workflow prioritization in radiology departments and serve as clinical decision support for non-imaging medical specialists, while opening the path for screening initiatives at population scale. \\\\
Computer vision is the interdisciplinary scientific field that seeks to automate the understanding of images by performing tasks that the human visual system can do \cite{Dana_Ballard1982-ge}. The success of Deep Learning (DL) algorithms for computer vision tasks, mainly with convolutional neural networks (CNNs), has led to a rapid adoption of these techniques in medical imaging research, boosting academic works that apply DL to diagnosis tasks, and promoting the building of large labeled medical imaging datasets that are needed to train these algorithms.   \\ \\
In the field of computer vision, different types of ground-truth labels are used to address different supervised tasks: classification CNNs are fed with image-level labels, object detection CNNs with bounding-boxes labels, and segmentation CNNs with pixel-level masks. Most large CXR datasets \cite{Wang2017-ql,Irvin2019-jy,Johnson2019-yp} have disease annotations in the form of positive or negative labels indicating the presence or absence of 14 findings that can appear in a CXR. Another large dataset released lately organizes labels as a hierarchical family of diseases \cite{Bustos2020-yv}. Following the release of these large CXR datasets, the use of DL for automated classification of CXR has been widely explored by the scientific community. Due to the type of labels, most research works applying CNNs for CXRs use classification architectures \cite{Wang2017-ql,Irvin2019-jy,Yao2017-hj,Rajpurkar2017-gm,Taylor2018-pa,Majkowska2020-vm,Annarumma2019-ga}, such as ResNet \cite{He2015-gb} or DenseNet \cite{Huang2016-qp}. However, the translation of DL tools for automated CXR interpretation to real clinical scenarios clearly faces challenges, as it is still poorly achieved in practice \cite{Liu2019-gc}. We believe this can be related to two aspects, which motivated this work. \\\\
In the first place, the difficulties that DL faces for CXR interpretation could be related to the multi-pathological approach of previous works. The use of DL in health has shown better results when applied to narrow tasks \cite{Allen2019-og} raising the concern that the detection of multiple diseases with the same underlying CNN model might be ineffective. Moreover, in clinical radiology CXR is considered a screening tool rather than a tool for differential diagnosis \cite{Kelly2012-wo}. It orients the diagnosis process and the choice of further medical studies, as a CXR study is usually insufficient to identify a specific disease with certainty. It remains one of the most complex imaging studies to interpret, being subject to significant inter-reader variability and suboptimal sensitivity for important clinical findings. Numerous pathologies are visually similar, which can lead to considerable variability between CXR reports, even among expert radiologists \cite{Neuman2012-de}. This reinforces the idea that multi-pathological CXR classification seems clinically inappropriate. To tackle this issue, image interpretation could be oriented to emulate the reading process of CXR by radiologists, recognizing radiological patterns and signs rather than individualizing multiple diseases \cite{Ellis2006-tc}. In medicine, the study of signs is referred to as semiology or semiotics. Radiological semiology refers to the description of particular imaging signs that can be observed and interpreted by an expert, and it plays a central role in imaging diagnosis \cite{Nordio2017-ut,Schiavon2008-fv}. In this regard, we propose a semiology-based detection, replacing the traditional 14-findings classification.\\\\
In the second place, image-level labels have no information on the localization of the finding in the image, and as most labels are assigned by automatic text mining of radiological reports (with no professional revision), they are subject to labeling errors. This might represent an unreliable ground-truth. Motivated by these limitations, some CXR datasets have been released lately containing stronger labels, such as bounding-boxes around pathological findings or pixel-level masks as regions of interest \cite{Shiraishi2000-ow,noauthor_undated-co,noauthor_undated-fe}. This enables the use of CNN architectures for object detection or segmentation.  To exploit both the localization information of strongly labeled datasets and the size of larger class-labeled datasets, we need algorithms that can be trained using all heterogeneous labels available from these different CXR datasets. The generalized adoption of classification architectures for detection of diseases in CXR seems limited in this matter.\\\\
In this work, we explore a solution that takes advantage of heterogeneous ground-truth labels from different CXR datasets. We present a simple approach that combines DL architectures for different computer vision tasks (image classification, object detection, and segmentation) as a late fusion of models, and provides a unified heatmap as an easily interpretable output for clinicians. The combined model will be referred to as TRx (named after the Spanish acronym for thoracic x-rays). We sought to develop a clinically-appropriate tool for computer-aided detection and triage of CXR findings, that recognizes the main radiological patterns relevant in diagnosis of CXRs, rather than differential diseases. The objectives of the present study were: (1) to report the model development, (2) to measure model performance in the local population.

\begin{figure*}[h]
  \includegraphics[width=\textwidth,keepaspectratio]{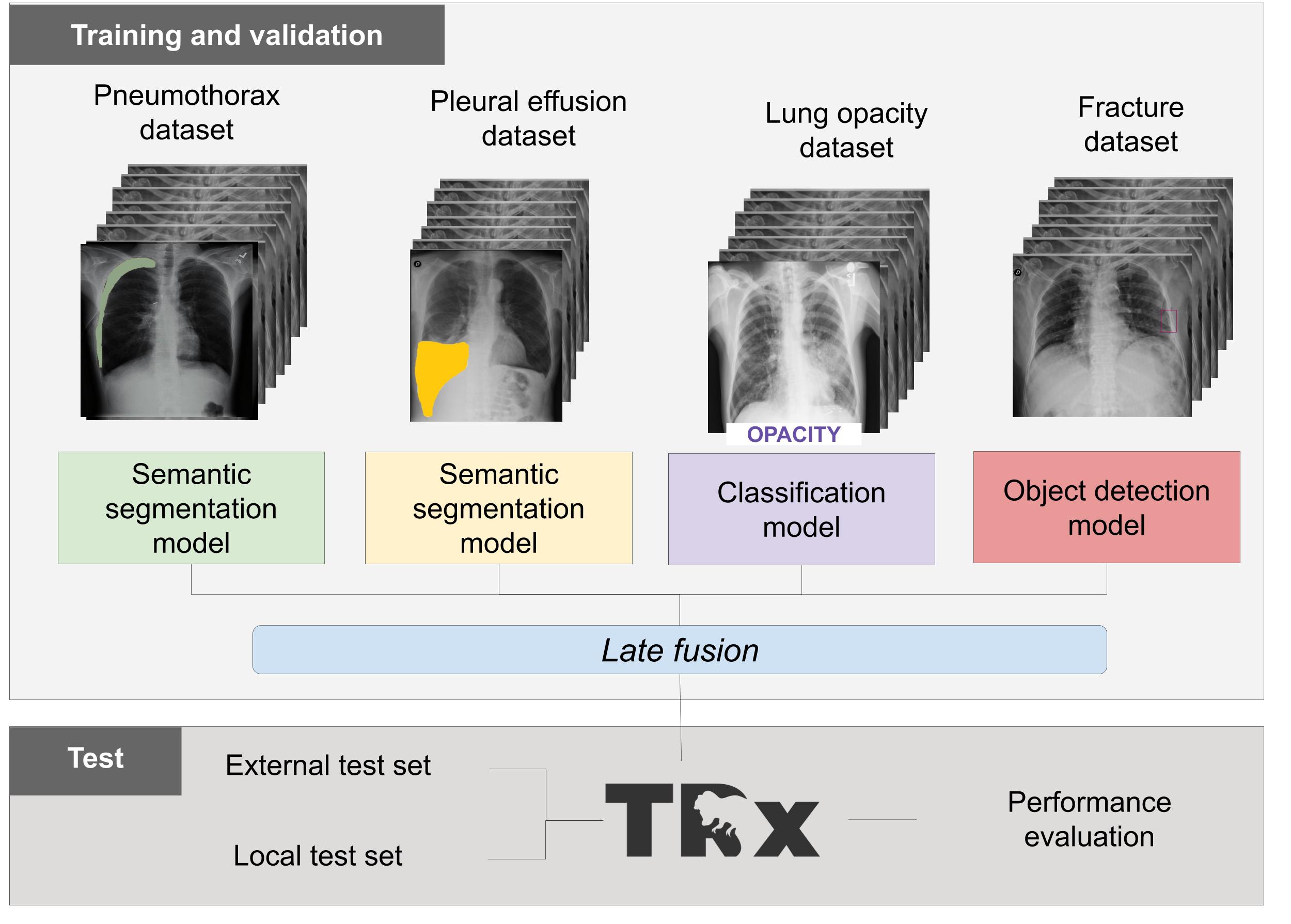}
  \centering
  %\fbox{\rule[-.5cm]{4cm}{4cm} \rule[-.5cm]{4cm}{0cm}}
  \caption{Development stages: organized in four independent training phases of Deep Learning models, and combined as one unified tool, named TRx.}
  \label{fig:fig1}
\end{figure*}
\section{Materials and Methods}
\subsection{Study design}
Local ethical review board committee approved the study, and waived informed consent given the anonymous and retrospective design. The intended use of this AI approach is as computer-aided detection and triage software for automatic detection of radiological findings in CXRs, with two main proposed clinical roles:
\begin{itemize}
\item For professionals on call at the Emergency Service, acting as an immediate diagnostic assistant, being a second opinion and helping professionals who may not have long experience in CXR interpretation, such as residents or non-imaging specialists. This is intended to integrate to our centre's Electronic Health Record application.
\item For automatic triage in the worklist for radiology reports of out-patients and emergency patients. In a Radiology Department, the end users are imaging specialists and therefore well-trained on CXR interpretation. These users might not find much added value on an automatic diagnosis-assistant tool. Rather, the intended use of TRx in this context is being an automatic triage tool, to improve resource allocation in CXR reporting and reduce the existing burden on this workflow, by assigning higher priority to those studies where pathological signs were detected by the DL model, so that they are reported sooner. 
\end{itemize}
The target radiological findings were chosen as those critically relevant for CXR triage interpretation. Taking the proposed clinical roles into account, this choice was based on a semiologic approach rather than a diagnosis-oriented approach. The four selected radiological findings were pneumothorax, fracture, pleural effusion, and lung opacity (which includes focal and diffuse opacities). \\\\
A central concept in this study was to use an appropriate type of ground-truth label for each of the four selected radiological findings. Based on the nature of each finding and following expert radiological opinion, the following scheme was defined: 
\begin{itemize}
\item \emph{Pleural effusion and pneumothorax}: annotated using pixel-level labels (masks). These findings manifest as well-defined areas in the lung, as they are caused by a volume of fluid, so the use of masks seems most appropriate.
\item \emph{Lung opacity}: annotated as image-level labels. In this case, we consider a general label is adequate as this finding is heterogeneous: it might be observed in an expanded manner across the lung tissue or in focalized areas. Pulmonary opacification, as a decrease in the ratio of air to soft tissue in the lung, may be caused by consolidation, ground-glass opacificacification, atelectasis or nodules as main differential diagnosis. 
\item \emph{Fracture}: annotated as bounding-box labels. This finding is well-localized but it is hard to label in a pixel-level manner, as the mask would be a thin strip through the fracture line. We consider that a fracture is better represented by a bounding-box.  \end{itemize}
The development process was divided into four independent modules, corresponding to four DL models, each focusing on the detection of one of the selected radiological findings. The chosen type of ground-truth label determined the choice of architecture for each model, which were trained with four independent training datasets. 
\newcolumntype{G}{>{\columncolor{gray!20!white}}p{0.1\textwidth}}
\begin{table}[h]
 \caption{Summary of the four Deep Learning modules.}
  \centering
 \begin{tabularx}{\linewidth}{G|XXXX} %p{0.2\textwidth}p{0.15\textwidth}p{0.15\textwidth}p{0.15\textwidth}}
    \toprule
    & Pneumothorax & Pleural Effusion & Lung opacity & Fracture \\ \midrule 
    Computer vision task & Image segmentation & Image segmentation & Image classification & Object detection \\ \hline
    Input & \multicolumn{4}{c}{ Gray images (one channel) } \\\hline
    CNN & AlbuNet-34 \cite{Shvets2018-qy} & AlbuNet-34 & Inception-ResnetV2 \cite{Szegedy2016-ht}* & RetinaNet \cite{Lin2017-bf} \\\hline
    Raw output & Mask image (an output score per pixel) & Mask image (an output score per pixel) & Two-class softmax (an output score per image) & List of bounding boxes (four coordinates and a confidence score per box) \\\hline
    Performance metric & Dice score & Dice score & Accuracy & Average Precision \\\hline
    Obtention of binary output & Threshold on the sum of all pixel scores & Threshold on the sum of all pixel scores & Threshold on the positive class softmax output & Threshold on the maximum score among detected boxes \\ \bottomrule
    \multicolumn{5}{l}{\footnotesize{CNN: Convolutional neural network.}}\\
    \multicolumn{5}{l}{\footnotesize{*Adapted to use one-channel inputs.}}\\
    
\end{tabularx}
  \label{tab:table1}
\end{table}
The final phase of the study was an external validation of the combined model performance using two test sets: a publicly available test set \cite{noauthor_undated-fl}, containing images from NIH ChestX-Ray14 dataset which were relabeled with a careful methodology by \cite{Majkowska2020-vm}; and a local retrospective collection of CXRs from our center, a 650-bed university hospital. This stage aimed to evaluate the four DL-models as one unified tool, named TRx, and generated as a late fusion of the models (Figure \ref{fig:fig1}).
\subsection{Data}
To build the training datasets we used a combination of images from public datasets and retrospective images from our center’s Picture Archiving and Communication System (PACS). We designed the building of four independent datasets. In each case, the source of images was chosen following a set of inclusion and exclusion selection rules, and the labeling strategy was designed following a specific rationale (details can be found in Supplementary Material A). Each training dataset was split into training and tuning subsets by a random partition at patient-level with a ratio 80:20, guaranteeing both a stratified split between images with findings and without findings, and no overlap of patients across different sets. Ground-truth annotation criteria are reported in Supplementary Material A. \\\\
Two test datasets were used. The external test set is a sample from NIH ChestX-ray14, relabeled and released by \cite{Majkowska2020-vm}. It was obtained from a public cloud repository \cite{noauthor_undated-fl}. As the four label categories of this dataset are slightly different than ours, we grouped them to allow comparison with TRx outputs, as follows: positive labels for air space opacity or nodule/mass were considered a positive label for lung opacity. No overlapping between training and test is guaranteed as we do not use any ChestX-ray14 images for none of the training datasets. \\\\
The local test set is a retrospective collection of de-identified images from our PACS, obtained by a search on radiological reports records of CXRs performed between 2008 and 2019 at a 650-bed university hospital. Inclusion criteria for CXR images were having a concluding report and belonging to a patient whose images were not used in any training dataset. A senior radiologist reviewed the images, together with the radiological report and patient records, to assign four binary labels indicating the presence or absence of each evaluated radiological finding. \\\\
A general abnormality label was built to classify images into two non-overlapping groups: positive label (abnormal) when at least one finding was present and negative label (normal) when no finding was present.   
\begin{figure*}[h]
  \includegraphics[width=\textwidth,keepaspectratio]{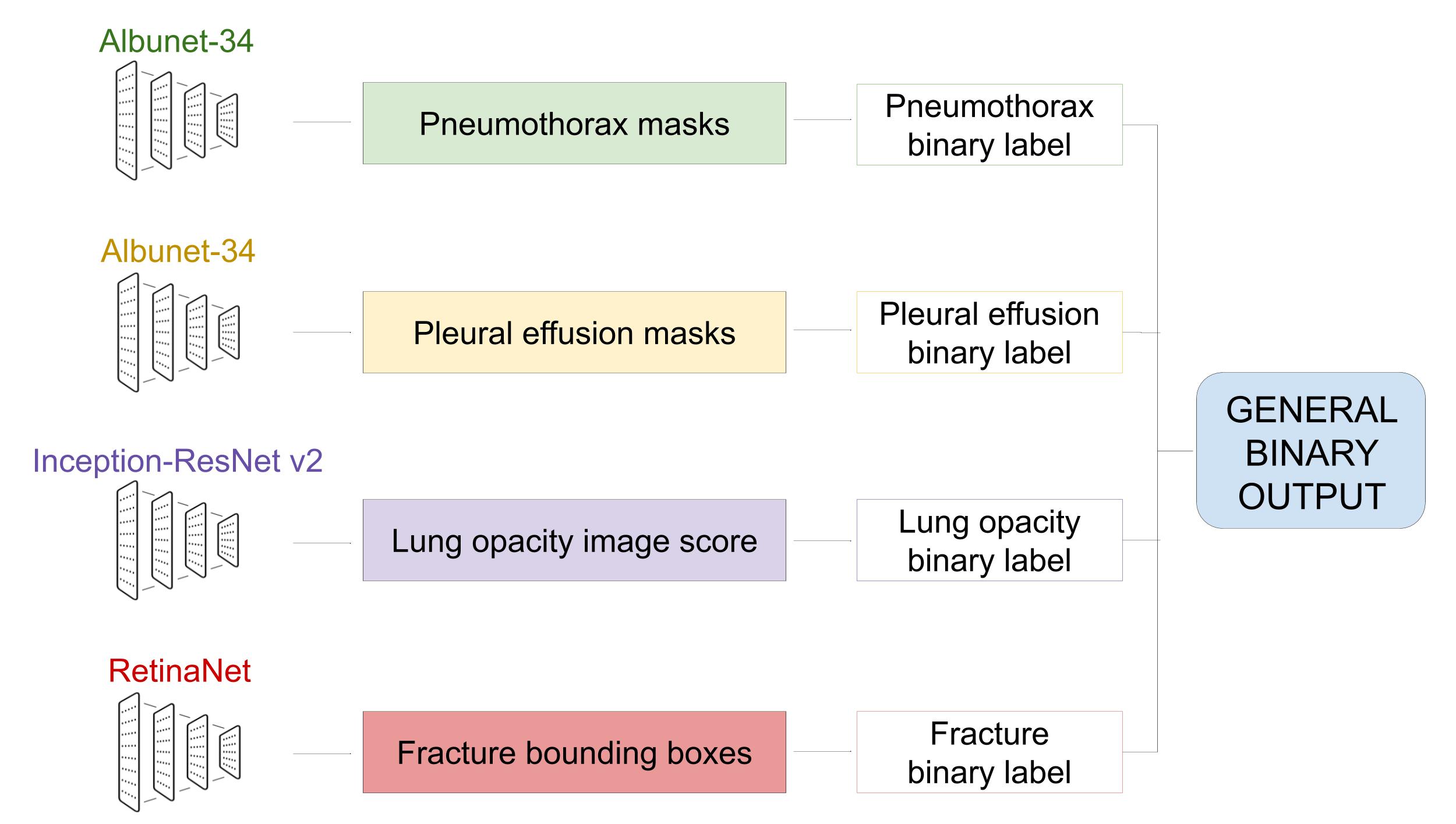}
  \centering
  %\fbox{\rule[-.5cm]{4cm}{4cm} \rule[-.5cm]{4cm}{0cm}}
  \caption{Pipeline designed to obtain the general binary output.}
  \label{fig:fig2}
\end{figure*}
\subsection{TRx: training and fusion strategy}
Four CNNs were trained, as indicated in Table \ref{tab:table1}. The details on data augmentation, implementation software and training hyperparameters are described in Supplementary Material B. In each case, the best model checkpoint was selected based on the corresponding performance metric, measured on the tuning set.\\\\
Late fusion was accomplished by first computing a binary prediction from each model output, and then calculating a general binary output as a logical ‘OR’ operator between these four binary outputs (if at least one is positive, the general output is positive), as shown in Figure \ref{fig:fig2}. The thresholding method used to binarize the output for each architecture is detailed in Table \ref{tab:table1} Cutting values were calibrated using the tuning set.  \\\\
To improve the interpretability of the final prediction, we developed a method to create one unique heatmap from the outputs of the four DL models. As each DL architecture has raw outputs of different nature, it is important to achieve an homogenous output that is easy to understand for clinicians. During inference of an image, the first step is to obtain a heatmap from each DL architecture’s output.  These four heatmaps are homogenized by using one common color scale. For the pneumothorax and pleural effusion models, each pixel color is determined by the linear output of a final 2D convolutional layer of unitary kernel size and one filter. For the lung opacity model we apply class activation maps obtained by global average pooling of the last convolutional layer’s output. For the fracture model, we built gradient color ellipses within each bounding box, with a red center that grows to blue borders. In all cases, non-activated regions are made transparent for ease of interpretation. Finally, the unified heatmap is obtained by overlapping these four individual heatmaps and applying a median blur filter of kernel size 5x5 to eliminate saturation noise.
\subsection{Evaluation/Statistic analysis}
We measured TRx diagnostic performance as a binary classification task: we used the general output of the late fusion of four models for detection of abnormal images. We also tested the performance of each individual model as four independent binary classification tasks. The main outcome variable for each binary test was the area under the Receiver Operating Curve (AUC). Sensitivity, specificity, positive predictive value, and negative predictive value were also measured. 95\% confidence intervals were obtained using 10,000 bootstrap samples. \\\\
Motivated by the future clinical implementation at our center, we performed an extra analysis for the local test set to identify possible population bias \cite{Larrazabal2020-xa}. The test was repeated for different demographic subgroups to evaluate fair performance \cite{Seyyed-Kalantari2020-ls}, as we consider it a critical aspect for clinical implementation. 
\section{Results}
\subsection{Datasets}
The four training datasets built for this study are summarized in Table \ref{tab:table2}. Relabeling was needed in three cases: we performed grouping of classes in the case of lung opacities, bounding-box annotation in the case of fractures, and segmentation in the case of pleural effusion. \\\\
The two test datasets are described in Table \ref{tab:table4}. The external test set contains 4376 frontal CXRs from 1695 patients, combining the validation and test sets released by \cite{Majkowska2020-vm}. The local test set contains 1064 frontal CXRs from 999 patients. Further demographic details on this institutional dataset can be found in Table C1.
\begin{table}[h]
\newcolumntype{g}{>{\columncolor{gray!20!white}}l}

 \caption{Description of training datasets.}
 \begin{tabularx}{\linewidth}{gg|XXXX}
    \toprule
    & & Pneumothorax & Pleural Effusion & Lung opacity & Fracture \\ \midrule
    \multicolumn{2}{g|}{ Images source } & Kaggle competition \cite{noauthor_undated-co} & Our center PACS & CheXpert \cite{Irvin2019-jy} & MIMIC-CXR \cite{Johnson2019-yp} \\ \hline
    \multicolumn{2}{g|}{ Relabeling } & None. & Manual segmentation of masks. & Automatic grouping of multicategorical labels. & Manual delineation of bounding boxes. \\ \hline
    \multicolumn{2}{g|}{ Number of patients } & 10,675 & 734 & 12,695 & 412 \\ \hline
    & Mean $\pm$ Std & 47$\pm$ 17 & 63 $\pm$ 19 & 56$\pm$ 18 & 59$\pm$ 18 \\ 
    \multirow{-2}{*}{Age (years) } & Median [IQR] & 49 [35-60] & 66 [52-78] & 57 [43-69] & 60 [49-72] \\ \hline
    & Female & 4,795 (45\%) & 341 (46\%) & 4,835 (38\%) & 248 (45\%) \\
    & Male  & 5,880 (55\%) & 380 (52\%) & 7,859 (62\%) & 304 (55\%) \\
     \multirow{-3}{*}{Sex }& Unknown/Other & - & 13 (2\%) & 1 & - \\ \hline
    \multicolumn{2}{g|}{ Number of images } & 10,675 & 891 & 15,826 & 554 \\ 
    \multirow{3}{*}{ } & With finding & 2,379 & 712 & 10,327 & 277 \\
    & Without finding & 8,296 & 179 & 5,499 & 277 \\
    & N° of findings & 2,379 & 712 & 10,327 & 509 \\
    \multicolumn{2}{g|}{\shortstack[l]{Number of images assigned \\as tuning set} } & 2,135 & 178 & 3165 & 109 \\
    \bottomrule
    \multicolumn{6}{l}{\footnotesize{PACS: Picture Archiving and Communication System.}}\\
    \multicolumn{6}{l}{\footnotesize{IQR: Inter quartile range. }}
\end{tabularx}
  \label{tab:table2}
\end{table}
\newcolumntype{g}{>{\columncolor{gray!20!white}}l}
\subsection{TRx performance}
\begin{figure*}[h]
  \includegraphics[width=\textwidth,keepaspectratio]{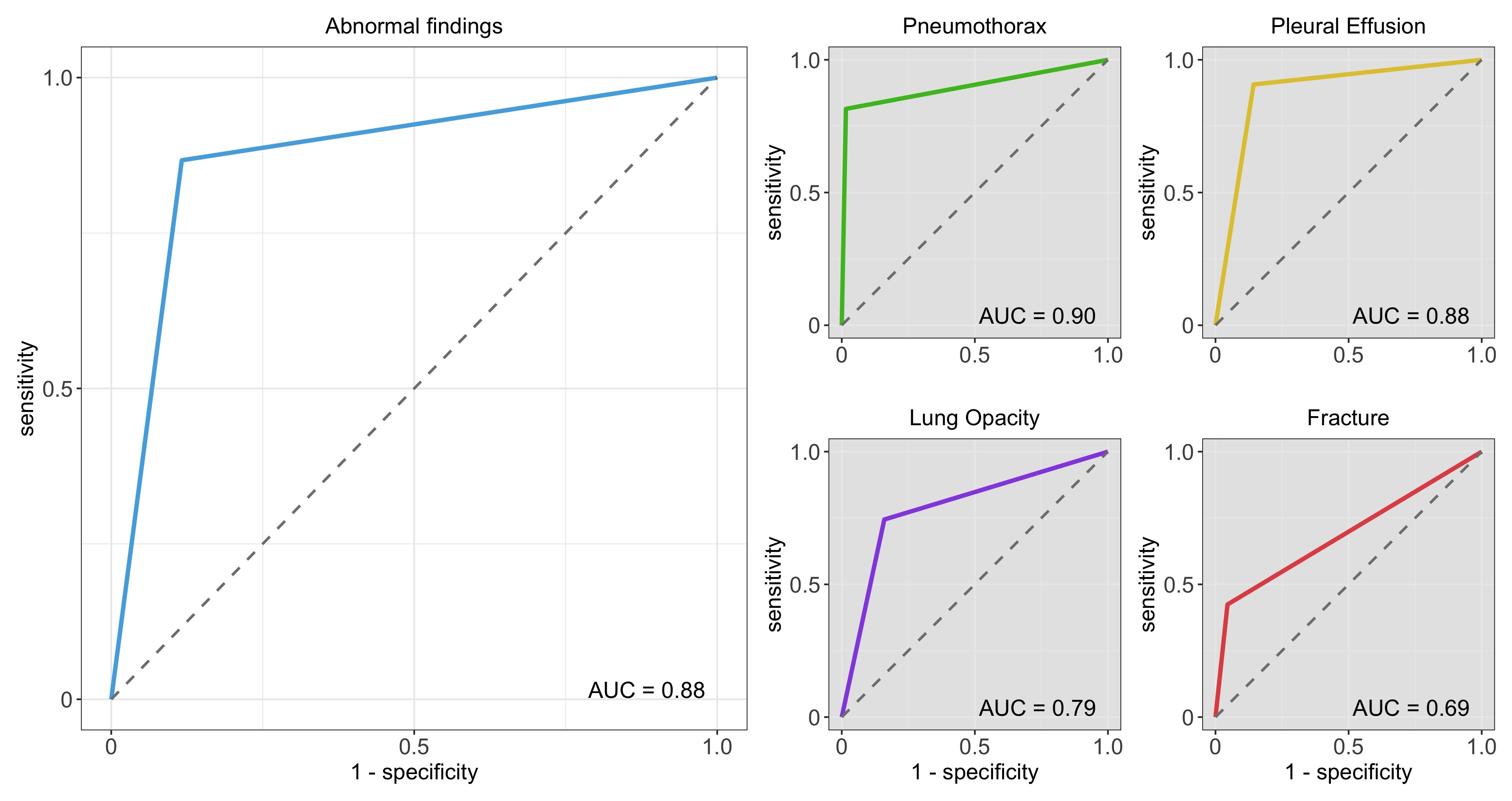}
  \centering
  %\fbox{\rule[-.5cm]{4cm}{4cm} \rule[-.5cm]{4cm}{0cm}}
  \caption{Receiver Operating Characteristics curves for the local test set.}
  \label{fig:fig3}
\end{figure*}

In abnormality detection, TRx presented an AUROC of 0.7491 (CI95: 0.74-0.76) and 0.8745 (CI95: 0.86-0.89) for the external and local test sets respectively. The best performance was obtained for pneumothorax detection, followed by pleural effusion, which correspond to the segmentation models. The worst performance was observed for fracture detection, corresponding to the smallest training dataset. This tendency is observed in both test sets. For pneumothorax detection, AUROC values were 0.89 (CI95:0.87-0.90) and 0.90 (CI95:0.86-0.89) for external and local sets respectively; for lung opacity these were 0.75 (CI95:0.74-0.76) and 0.79 (CI95: 0.77-0.82), and  for fracture 0.59 (CI95:0.56-0.92) and 0.69 (CI95:0.64-0.74). Pleural effusion was only measured in the local dataset, as the external dataset does not include this label class, presenting an AUROC of 0.88 (CI95:0.85-0.91). These values and their confidence intervals can be found in Table C2. 
\begin{table}[h]
 \caption{Description of test datasets.}
 \begin{tabularx}{\linewidth}{gg|XX}
    \toprule
    & & External test set & Local test set \\ \midrule
    \multicolumn{2}{g|}{ Images source } & Chest X-Ray14 \cite{noauthor_undated-fl} & Our center’s PACS \\ \hline
    \multicolumn{2}{g|}{ Number of patients } & 1695 & 999 \\ \hline
    & Mean $\pm$  Std & 47 $\pm$  16 & 58 $\pm$ 21 \\
    \multirow{-2}{*}{Age (y) } & Median [IQR] & 49 [34-58] & 48 [32-64] \\ \hline
     & Female & 1765 (60\%) & 543 (51\%) \\
    \multirow{-2}{*}{Sex }& Male& 2611 (40\%) & 521(49\%) \\ \hline
    \multicolumn{2}{g|}{ Number of images } & 4376 & 1064 \\ \hline
    \multicolumn{2}{g|}{ Normal
    (no finding) } & 1877(57\%) & 703 (66\%) \\
    \multicolumn{2}{g|}{ Abnormal
    (at least one finding) } & 2499(43\%) & 361 (34\%) \\
    & Pneumothorax & 238 & 119 \\
    & Pleural effusion & - & 86 \\
    & Lung opacity & 2373* & 176 \\
    & Fracture & 186 & 66 \\
    \bottomrule
    \multicolumn{4}{l}{\footnotesize{PACS: Picture Archiving and Communication System.}}\\
    \multicolumn{4}{l}{\footnotesize{IQR: Inter quartile range. }}\\
    \multicolumn{4}{l}{\footnotesize{*Lung opacity is considered present when at least one of Airspace opacity or Nodule/mass original labels are positive}}
\end{tabularx}
  \label{tab:table3}
\end{table}
Figure \ref{fig:fig3} shows the ROC curves for each binary classification task, measured in the local test set. Table \ref{tab:table4} presents the main diagnosis metrics obtained for this set. Performance analysis across demographic subgroups is included in Supplementary Material C. No evidence of population bias was observed.
\newcolumntype{g}{>{\columncolor{gray!20!white}}l}
\begin{table}[h]

 \caption{Diagnosis metrics for the local test set.}
 \begin{tabularx}{\linewidth}{g|XXXXX}
 
    \toprule
& Abnormality & Pneumothorax & Pleural effusion & Lung opacity & Fracture \\ \midrule
Sensitivity & \shortstack{86.61$\%$ \\
\small [83.62-89.61]*} &\shortstack{ 81.71$\%$\\
\small[75.22-87.39] }& \shortstack{90.82$\%$ \\
\small[85.33-95.56]} & \shortstack{74.48$\%$\\
\small[69.19-79.56]} &\shortstack{ 42.24$\%$\\
\small[31.34-52.70]} \\ \hline
Specificity & \shortstack{88.29$\%$\\
\small[86.29-90.23]} & \shortstack{98.40$\%$\\
\small[97.66-99.05]} & \shortstack{85.69$\%$\\
\small[83.87-87.40]} &\shortstack{ 84.04$\%$\\
\small[81.94-86.14]} & \shortstack{95.48$\%$\\
\small[94.36-96.59]} \\ \hline
\shortstack[l]{Positive \\predictive value} &\shortstack{ 79.13$\%$\\
\small[75.84-82.54] }&\shortstack{ 86.51$\%$\\
\small[80.80-91.80]} & \shortstack{35.93$\%$ \\
\small[30.77-41.31] }&\shortstack{ 48.04$\%$\\
\small[43.08-53.36] }& \shortstack{38.26$\%$\\
\small[28.57-48.61] }\\ \hline
\shortstack[l]{Negative \\predictive value} & \shortstack{92.78$\%$\\
\small[91.06-94.40] }&\shortstack{ 97.72$\%$\\
\small[96.84-98.52]} &\shortstack{ 99.06$\%$\\
\small[98.46-99.54]} & \shortstack{94.33$\%$\\
\small[93.00-95.65]} & \shortstack{96.14$\%$\\
\small[95.905-97.17]} \\
\bottomrule
\multicolumn{6}{l}{\footnotesize{*Values in brackets correspond to 95$\%$ confidence intervals.
}}
\end{tabularx}
  \label{tab:table4}
\end{table}
\subsection{Detection examples}
\begin{figure*}
  \includegraphics[width=\textwidth,keepaspectratio]{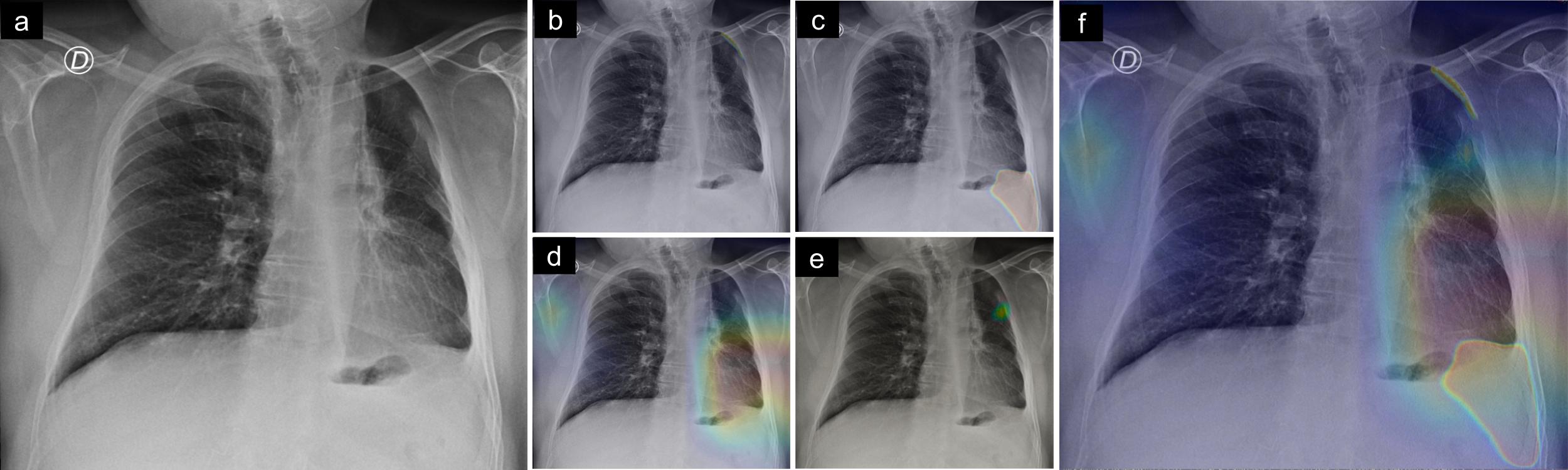}
  \centering
  %\fbox{\rule[-.5cm]{4cm}{4cm} \rule[-.5cm]{4cm}{0cm}}
  \caption{(a) Original image: early post-operative chest x-ray following left upper lobectomy with residual pneumothorax. (b) Pneumothorax finding at left top lobule (c) Pleural effusion finding at left bottom lobule. (d) Lung opacity class activation map (finding at left bottom lobule). (e) Fracture heatmap, built from bounding box. (f) Unified heatmap combining all four individual-finding heatmaps.}
  \label{fig:fig4}
\end{figure*}

The main visual output of the model is the unified heatmap, and the four individual-finding heatmaps are secondary outputs which are also available for clinician interpretation. Figure \ref{fig:fig4} illustrates an example, showing the original CXR, its corresponding individual-finding heatmaps and the unified heatmap. 
\begin{figure*}
  \includegraphics[width=\textwidth,keepaspectratio]{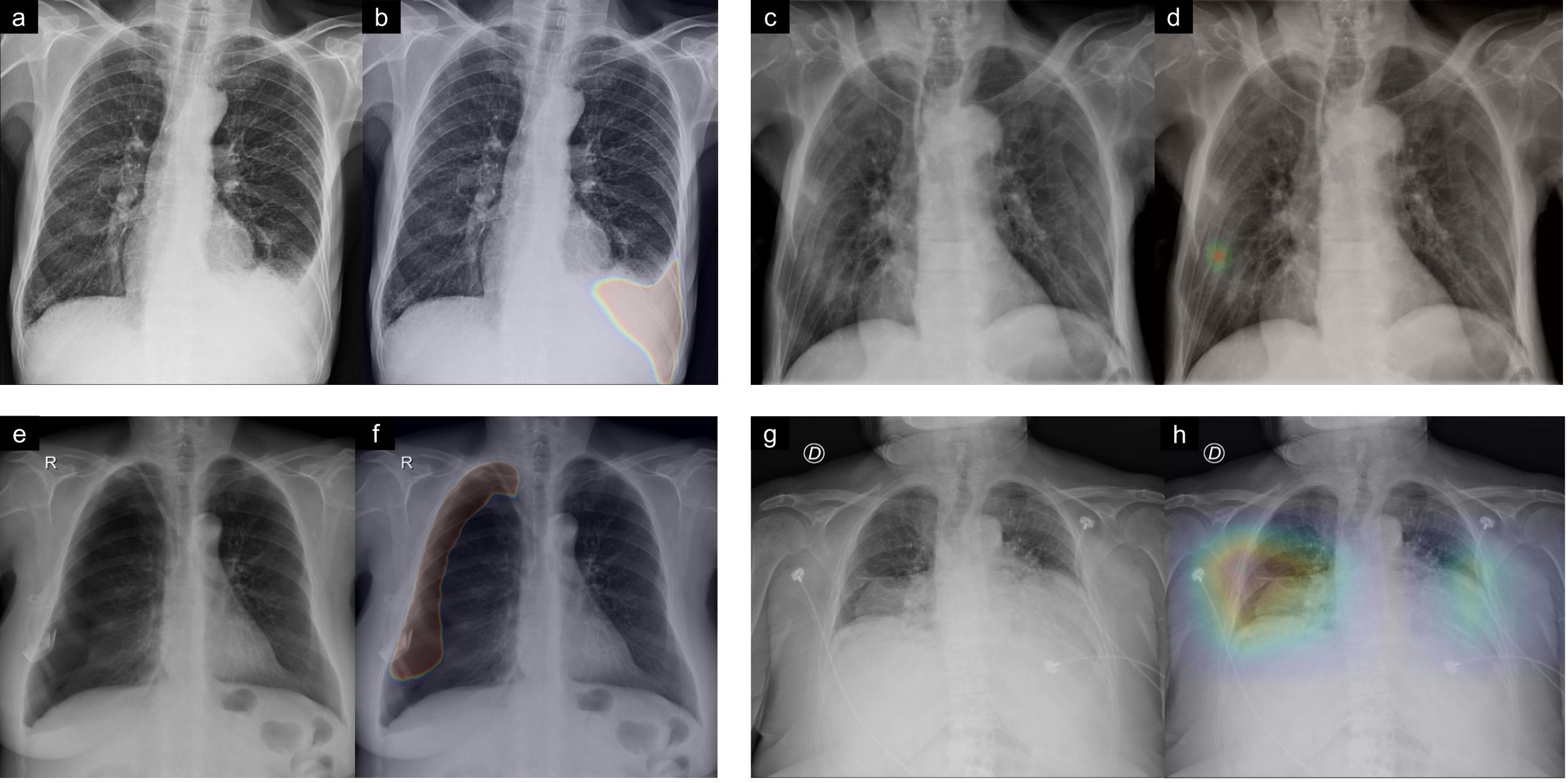}
  \centering
  %\fbox{\rule[-.5cm]{4cm}{4cm} \rule[-.5cm]{4cm}{0cm}}
  \caption{True positive examples. (a) Chest x-ray presenting mild left lower pleural effusion. (b) TRx detection of regions affected by pleural effusion. (c) Chest x-ray with multiple rib fractures in the right lower rib cage. (d) TRx localized detection for fracture. (e) Large pneumothorax in right lung. (f) TRx detected area covers almost all pneumothorax. (g) Chest x-ray presenting lung opacities in both lower pulmonary fields. (h) TRx lung-opacity heatmap, activated in both lungs.
}
  \label{fig:fig5}
\end{figure*}
Figure \ref{fig:fig5} shows examples of true positives for each individual heatmap: cases where TRx correctly detected the radiological finding. Figure \ref{fig:fig6} shows cases of false positives, where TRx wrongly detected a finding that is not present or cannot be confirmed. They are cases of typical confusers for radiological CXR findings, where misinterpretations are expected: calcifications mistaken for fractures, apical bullae confused with pneumothorax, emphysemas wrongly interpreted as effusion, or external devices such as prosthesis causing non-pathological opacities. These use cases highlight the importance of clinicians in the diagnosis loop, showing why interpretable outputs are necessary and why the examination by a medical specialist is essential. The unified heatmaps for these cases and some other examples are depicted in Figure C4 in the Supplementary Material.
\begin{figure*}
  \includegraphics[width=\textwidth,keepaspectratio]{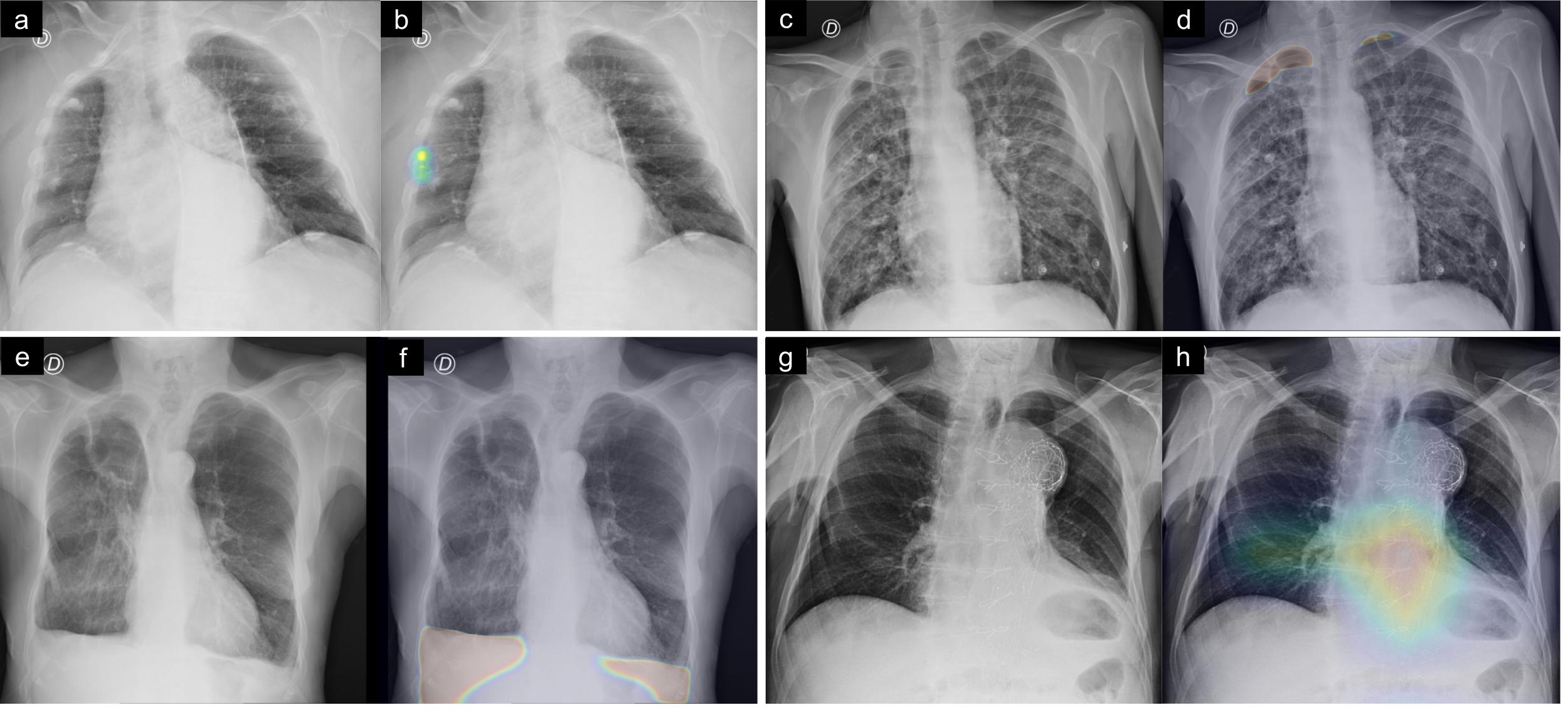}
  \centering
  %\fbox{\rule[-.5cm]{4cm}{4cm} \rule[-.5cm]{4cm}{0cm}}
  \caption{False positives examples. Pairs of images are the original CXR (left) and the individual finding heatmap being evaluated (right). (a,b) Pleural calcifications in the right middle rib cage are misinterpreted as a fracture. (c,d) Subpleural apical bullae is confused with pneumothorax. (e,f) Emphysema in both lower lungs causes blunting of the left costophrenic angle, which is incorrectly detected as pleural effusion. (g,h) Aortic prosthesis causes false positive detection of lung opacities.}
  \label{fig:fig6}
\end{figure*}
\section{Discussion}
This study presents a new approach for exploiting heterogeneous labels from different CXR datasets, by choosing CNN architectures according to the semiology of each pathological finding. The objective is not to build a tool for differential diagnosis but to detect imaging signs and patterns that can appear in a CXR. We believe this approach is closer to the actual reading process of CXRs by professionals, and therefore more likely to be successful in a real clinical setting.\\ \\
Extensive work has been performed using large CXR datasets with image-level labels to train DL classification models. Wang et al. \cite{Wang2017-ql} compared four widely-used architectures (AlexNet, GoogLeNet, VGGNet-16 and ResNet-50) on the ChestX-Ray14 dataset, with an average AUROC across pathologies of 0.73 for the best performing model on the test set. Rajpurkar et al. \cite{Rajpurkar2017-gm} reported an average test AUROC of 0.84 using DenseNet architecture for this dataset, while Guan et al. \cite{Guan2018-ol} obtained 0.87 applying an attention-guided model, and Guendel et al. \cite{Guendel2019-jv} reported 0.88 using a multi-task architecture that supports segmentation of lungs and heart. On the CheXpert dataset, Irvin et al. \cite{Irvin2019-jy} achieved an average AUROC of 0.908 across five selected pathologies. Bressem et al. \cite{Bressem2020-gk} performed a systematic comparison of classification architectures for the CheXpert dataset, reporting an average AUROC of 0.88. On the MIMIC-CXR dataset, Rubin et al. \cite{Rubin2018-jw} reported an average AUROC of 0.721 across 14 findings when using a dual-input DenseNet-based model trained with both frontal and lateral CXRs. Other works exploited both the image-level annotationes and the available bounding box annotations in the ChestX-Ray14 dataset, to simultaneously perform disease identification and localization through the same underlying model \cite{Li2017-hx,Rozenberg2020-kh} with a multiple instance learning algorithm. They reported that it requires less image-level annotated images to achieve similar AUROC scores by combining them with a small set of bounding-box annotated images for training. Moreover, works that use strongly labeled datasets (such as RSNA Pneumonia or SIIM-ACR Pneumothorax) usually report a localization metric such as DICE rather than an AUROC value \cite{Gabruseva2020-qt,Groza2020-kf,Tolkachev2020-mp,Khan2020-se}. To our knowledge, no works have yet reported using multiple CXR datasets with heterogeneous annotations. \\\\
An important challenge when translating DL models to clinical workflow is the change in performance caused by cross-domain. The limitations on cross-domain for CXR automatic diagnosis across different datasets have been explored by \cite{Cohen2020-gk}, who characterized how performance metrics are generally not translated to new CXR domains. Most works mentioned above use train and test sets that are obtained by splitting the same original dataset, avoiding this particular challenge. Moreover, as it cannot be guaranteed that a performance level observed with public test datasets will be consistently maintained in our local clinical workflow, we evaluated our model in a set of local images. We obtained an average AUROC of 0.82 across the four selected findings, and an abnormality detection AUROC of 0.88. Although test sets and target findings differ among related works, making it difficult to compare results straightforwardly, these values are competitive with state-of-the-art results, and represent a closer estimation of what performance might be if implemented in a real-world scenario at out center. \\\\
In an attempt to obtain reproducible metrics for comparison with related works, we evaluated TRx on an external publicly available dataset, whose ground-truth labels match closer to our four findings approach. This dataset is part of the work from \cite{Majkowska2020-vm}, who used a combination of a locally-built dataset and a relabeling of the ChestX-Ray14 dataset with a carefully-designed strategy for ground-truth assignment. Their total number of training images was 657,954, with around 5\% of images manually labeled by professionals and the remaining annotated by automatic text labelers applied to radiological reports. They obtained an average AUROC of 0.86 on their local test set and 0.90 on the relabeled ChestX-Ray14 test set across four findings, using four deep learning ensembles of several checkpoints each. We evaluated our model on a combination of their validation and test relabeled sets. As our model’s labels differed from this test set’s labels, we merged categories both in ground-truth and in predictions. We obtained a mean AUROC of 0.74 on this combined set of 4376 images, and this value was maintained when testing only on their test set of 1962 images. Comparison can therefore be made between the AUROC of 0.90 obtained by the authors and TRx AUROC of 0.74. The merging of label classes could explain this lower AUROC observed.\\\\
Our strategy was a late fusion of four independent CNNs. This allowed us to take advantage of multiple publicly available datasets, to focus on the most appropriate computer vision task for each radiological finding, and to give priority to datasets with manually assigned annotation. We believe the simplicity of this approach is a strength that could be further exploited, and might be useful in tackling the limits of cross-domain generalization in CXR diagnosis as it combines many sources of images. Moreover, using independent architectures for each radiological finding enables the design of individualized quantification methods and descriptive approaches that are particularly relevant for each disease (for example, lung area affected is a useful metric for effusion, whereas fractures can be better described by counting the number of fracture instances). \\\\
A limitation of this work is ignoring the relationships and dependencies that exist between diseases. New architectures and loss functions that exploit heterogeneous labels through the same underlying model could be explored, allowing the CNN to learn these relationships. Moreover, we excluded lateral CXR images, which might improve performance, particularly for findings such as small pleural effusions that might not be visible in frontal view.\\\\
This work can be considered as an exploratory study, focused on translation to clinical scenarios, that addresses the use of heterogeneous labels in one unified DL tool. Besides reaching state-of-the-art performance metrics, we sought to address other important aspects of clinical implementation in public institutions. Firstly, we kept a low computational cost, which makes integration feasible. While other works use up to 30 model checkpoints \cite{Majkowska2020-vm}, we use only four, reducing the computational power needed on processing servers. Secondly, we developed a friendly interpretable output, which is a unified heatmap and one binary output. This attempts to achieve a successful adoption of TRx as CADe by clinicians and imaging specialists. Another important step before clinical integration is the validation in a local environment, for which we performed a retrospective test, with an evaluation of fair performance among demographic subgroups \cite{Seyyed-Kalantari2020-ls}. We analyzed the demographic variables that were available in patient information, limited to gender and age, finding no significant difference among groups. Future work includes performing a prospective validation both at our center and at external centers. In these validation experiences, comparison of performance should not be limited to imaging specialists, including professionals from other medical specialties as well, such as emergency or family physicians, who are not specifically trained in CXR reading and could further benefit from automatic assistance. 
\section{Conclusion}
The semiologic approach and the late fusion strategy introduced in this work  showed promising results in both external and local populations. We believe it is necessary to further explore the use of heterogeneous labels in CXR interpretation, while working with clinically sensible definitions for study design, in order to exploit the available annotated information in the most adequate way. Regarding TRx, the next step will be a clinical validation in a prospective set of images, with careful ground-truth assignment.  
\section{Acknowledgements}
We thank the imaging professionals from the Radiology Department at Hospital Italiano de Buenos Aires for their contribution with expert opinion. Additionally, we acknowledge the collaboration of all team members participating in the Program for Artificial Intelligence in Health at this hospital.\\\\
This work was supported by the annual research grant provided by Hospital Italiano de Buenos Aires. The Titan V used for this research was donated by the NVIDIA Corporation.

\bibliographystyle{unsrt}  
\bibliography{ms}  %%% Remove comment to use the external .bib file (using bibtex).

%%\begin{thebibliography}{1}

%%\end{thebibliography}

\end{document}

\begin{center}
  \textbf{ Chest x-ray automated triage: a semiologic approach designed for clinical implementation, exploiting different types of labels through a combination of four Deep Learning architectures}\\[1cm]\textbf{\large  Supplementary Material}\\[1cm]
 Candelaria Mosquera,\textsuperscript{1,2,*}
Facundo Diaz,\textsuperscript{3}
Fernando Binder,\textsuperscript{1}
José Martín Rabellino,\textsuperscript{3}\\
 Sonia Elizabeth Benitez,\textsuperscript{1}
 Alejandro Beresñak,\textsuperscript{3}
 Alberto Seehaus,\textsuperscript{3}
Gabriel Ducrey,\textsuperscript{3}\\
Jorge Alberto Ocantos,\textsuperscript{3}
 Daniel Roberto Luna,\textsuperscript{1}\\[0.5cm]
\textsuperscript{1}{Health Informatics Department, Hospital Italiano de Buenos Aires, Argentina}\\
\textsuperscript{2}{Universidad Tecnológica Nacional, Argentina}\\
\textsuperscript{3}{Radiology Department, Hospital Italiano de Buenos Aires, Argentina}\\
\textsuperscript{*}{\emph{Correspondence author: candelaria.mosquera@hospitalitaliano.org.ar. Juan Domingo Perón 4190}, \emph{C1199AAB, Ciudad Autónoma de Buenos Aires, Argentina. Tel (+54) 01149590200 - Int 505}}
(Dated: \today)\\[1cm]
\end{center}

\section*{A. Datasets}
\renewcommand{\thefigure}{A\arabic{figure}}
\renewcommand{\thetable}{A\arabic{figure}}

\subsection*{Building of training datasets}
\begin{figure*}[h]
  \includegraphics[width=\textwidth,keepaspectratio]{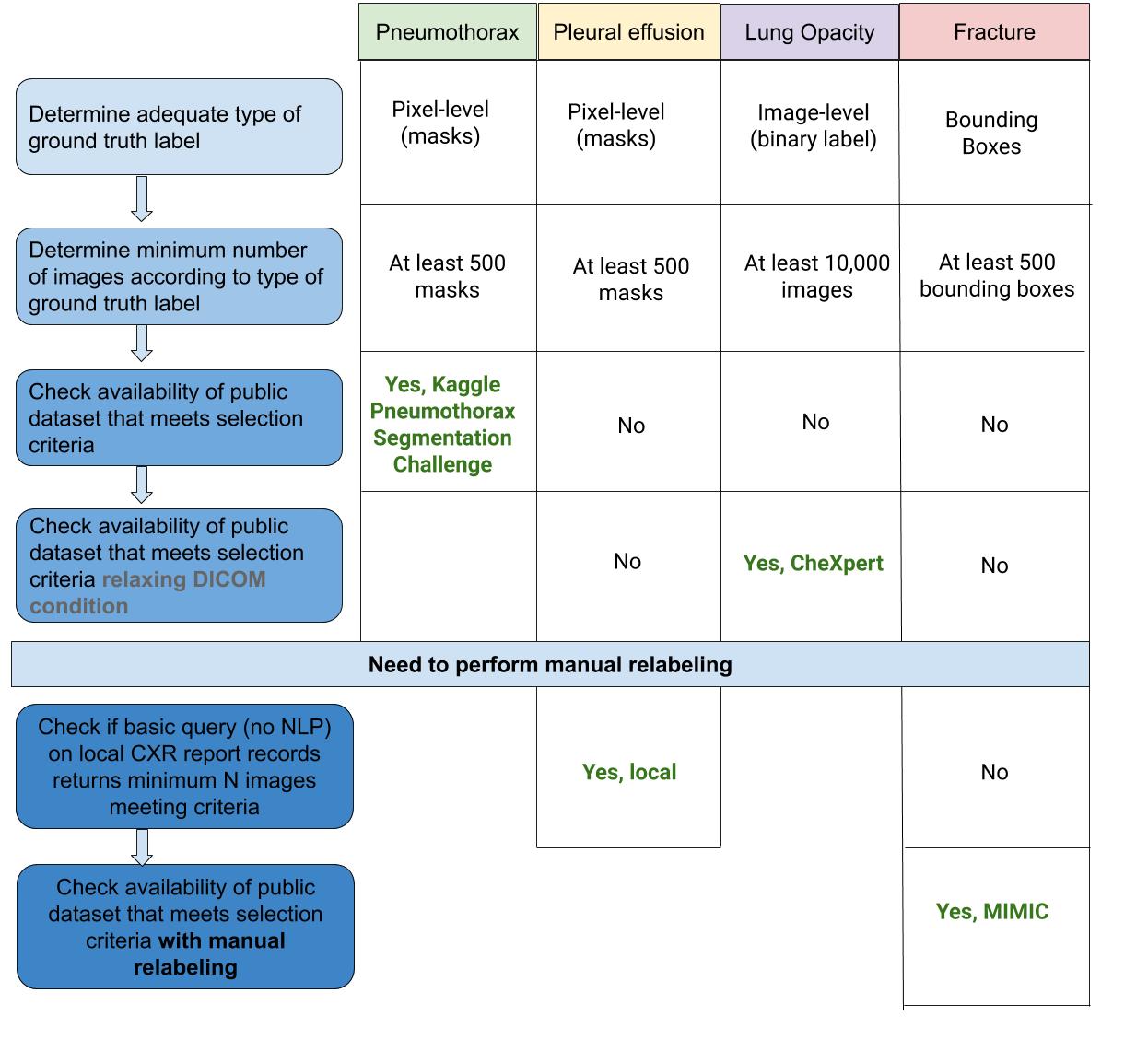}
  \centering
  %\fbox{\rule[-.5cm]{4cm}{4cm} \rule[-.5cm]{4cm}{0cm}}
  \caption{Selection criteria for choosing image sources. NLP: Natural Language Processing. CXR: chest x-ray.}
  \label{fig:figA1}
\end{figure*}

We chose datasets which had at least the minimum required number of images that meet the selection criteria. This criteria was being frontal chest x-rays (CXRs), in DICOM format, and having a ground-truth label either manually assigned (in the case of masks and bounding boxes) or automatically assigned by a method with F1 score greater than 0.90 as reported quality.  \\\
Regarding view position, it is well-known that anteroposterior patient orientation during CXR acquisition results in images with magnifications of vessels and mediastinum. As this affects particularly the detection of lung opacities, for this training set to be representative of the intended-use population (namely CXRs acquired from emergency or out-patients), we excluded anteroposterior images from this dataset.  \\
For each radiological finding, we built a dataset following the steps described below:
\begin{enumerate}

\item Determine the type of ground-truth label that is adequate for imaging characteristics of the target finding. 
\item Determine minimum number of images required in the training set. 
\item Check availability of a public data source that meets criteria
\item If no adequate data source is available, manual relabeling needs to be performed. Check if a basic query on radiological reports records from our center (standard searches, with no natural language processing) returns the minimum number of images needed.
\item If there is this number of local images available, perform relabeling on those images until reaching the minimum number needed.
\item Otherwise, check if there is a public data source which would meet selection criteria if relabeled. If so, relabel at least the minimum number of images needed.
\end{enumerate}
The results of applying these steps to each module are shown in Figure \ref{fig:figA1}.
\subsection*{Data sources descriptions}
The CXR images used in this study were obtained from a variety of data sources. Three public datasets were used, as well as two datasets built from local Picture Archiving and Communication System (PACS).  \begin{itemize}
    \item \textbf{CheXpert from Stanford Hospital} \cite{Irvin2019-jy} contains 224,316 images from 65,240 patients, labeled for the presence of 14 observations as positive, negative, or uncertain. Labels were assigned by natural language processing of radiology reports. The dataset includes a training and a validation subset, which we joined into one unique dataset. 
\item \textbf{Kaggle SIIM-ACR Pneumothorax Segmentation Challenge} \cite{noauthor_undated-fe} contains 10,712 chest x-ray images collected by the Society for Imaging Informatics in Medicine in collaboration with the American College of Radiology and the Society of Thoracic Radiology. Images were annotated using a free platform, indicating a binary label for the presence or absence of pneumothorax, and creating binary masks over the affected lung area. The dataset contains 8,333 images with no pneumothorax and 2,379 with pneumothorax. 
\item \textbf{MIMIC Chest X-ray Database} \cite{Johnson2019-yp} contains 377,110 chest radiographs in DICOM format corresponding to 227,835 radiographic studies performed at the Beth Israel Deaconess Medical Center, with structured labels derived from free-text radiology reports.
\item \textbf{A pleural effusion dataset} was built from our center imaging archive. A search was performed over radiological reports from the period between 2008 and 2018 to identify CXR reports containing reference to a pleural effusion finding. Inclusion criteria included being from an adult patient and acquired either at the Emergency service or from an out-patient. Hospitalized patients were excluded. The most recent 1000 studies were chosen, obtaining a final number of 891 images, which were de-identified and segmented to obtain pleural effusion masks. 
\item \textbf{Local test set} was built from our centre PACS. A search was performed over radiological reports from the period between 2008 and 2018 to identify CXR reports containing reference to any of the four findings evaluated in this study. Images already selected for the pleural effusion dataset were excluded. 
\item \textbf{External test set} was built by joining the final validation set and final test set from ChestX-ray14 \cite{noauthor_undated-fl} sampled and relabeled by \cite{Majkowska2020-vm}.
\end{itemize}

\subsection*{Ground-truth labeling }
\begin{itemize}
\item \textbf{Lung opacity training set:} CheXpert dataset has fourteen label classes, corresponding to fourteen different findings, which are non mutually-exclusive. Each label can be one of the following four types: confirmed positive (1), confirmed negative (0), uncertain (-1), or empty. We excluded images with a confirmed “Support Devices” label, as these findings can result in noisy datasets. All lateral images were excluded. A general “opacity” label was created, which unifies the following labels: Atelectasis, Edema, Consolidation, Pneumonia and Lung Opacity. The unifying criteria consisted in assigning a positive opacity label to an image when at least one of these five categories had a confirmed positive label. Otherwise, if at least one of these categories had an “uncertain” label, the opacity label was set as uncertain. Finally, if all five labels were either confirmed negative or empty, opacity label was set as confirmed negative. We kept images that verified one of the following two conditions:
\begin{itemize}
\item No Finding category with positive label or uncertain label (images with no findings)
 \item Opacity category with positive label or uncertain label (images with lung opacity)
 \end{itemize}
Labels were set as continuous values: confirmed positive as 0.99, confirmed negative or empty as 0.01, and uncertain as 0.6. Using categorical cross entropy as loss function, these values yield a greater penalty for uncertain images classified as negative than for uncertain images classified as positive. These values can work as a quantitative representation of the uncertainty of labels assigned by natural language processing.
\item \textbf{Pneumothorax training set:} no relabeling was needed. Mask information is available in the Kaggle competition, informed with run-length encoding. The details on the labeling process are detailed in the Kaggle competition information \cite{noauthor_undated-fe}.
\item \textbf{Pleural effusion training set:} an imaging specialist with five years of experience delineated polygon masks on lung areas affected by pleural fluid. Masks were verified by an imaging specialist of ten years of experience. The labeling work was done using Amazon Web Service, with SageMaker Labeling Jobs tool.
\item \textbf{Fracture training set:} an imaging specialist with five years of experience delineated bounding boxes over fractures. Annotations were verified by an imaging specialist of ten years of experience. The labeling work was done using Amazon Web Service, with SageMaker Labeling Jobs tool. 
\item \textbf{Local test set:} ground-truth was built from reviewing all study and patient information available, including original radiological report, and with visual confirmation of the finding in the CXR image. This process was performed by an imaging specialist with five years of experience.
\end{itemize}

\section*{B. Training details}
\renewcommand{\thefigure}{B\arabic{figure}}
\renewcommand{\thetable}{B\arabic{figure}}

Training was performed in four different phases. In all cases, the available dataset was divided into training and tuning sets, using a stratified split between images with and without findings, assigning 80\% of the dataset to the train set and 20\% to the tuning set. Splitting was performed at patient level, guaranteeing no patient is repeated across sets. Selection of training hyperparameters was done choosing the combination of hyperparameters that showed the best performance metrics on the tuning set. For each module, the threshold values used to convert the raw outputs into one binary output per image was chosen as the cutting point that maximized the area under the receiver operating characteristics curve (AUROC) on the tuning set, considering a binary classification task. Model training was performed with an NVIDIA GPU Titan V.\\
In this section we describe the training procedures followed in each phase. Pneumothorax and pleural effusion modules were trained using analogous procedures, as both use masks as ground-truth.

\subsection*{AlbuNet-34: Detection of pneumothorax and pleural effusion}

This architecture, published by Shvets et al. \cite{Shvets2018-qy}, was implemented in Pytorch, with a random weight initialization. Input image size was 1024x0124, in batches of two images, using Adam optimizer. As loss function we used a weighted sum of three different loss functions: 
$$ Loss = 3(\mbox{Binary Cross Entropy Loss}) + 4(\mbox{Focal Loss}) + (\mbox{Dice Loss})$$
Training of AlbuNet-34 was performed in four sequential schemes, varying learning rate (LR) strategy and varying the proportion of images with finding. 
\begin{enumerate}
    \item Initial learning rate of 1e-3, applying LR reduction on plateau by a factor of 10\% and with two epochs of patience, and sampling images as 90% with finding and 10\% without finding.
\item Initial learning rate of 1e-5, applying cosine annealing to LR, and sampling images as 60\% with finding and 40\% without finding.
\item Initial learning rate of 1e-5, applying cosine annealing to LR, and sampling images as 40\% with finding and 60\% without finding.
\item Initial learning rate of 1e-3, applying LR reduction on plateau by a factor of 10\% and with two epochs of patience, and sampling images as 30\% with finding and 70\% without finding.
\end{enumerate}

Data augmentation was performed using Albumentations package, in a compose transformation applying random combinations of the following transforms:

\begin{itemize}
    \item RandomContrast between -0.2 and 0.2
\item RandomGamma between 80 and 120
\item RandomBrightness between -0.2 and 0.2
\item ElasticTransform 
\item GridDistortion 
\item  OpticalDistortion 
\item Shift of -0.0625 and 0.0625 of image size
\item Scale between -0.1 and 0.1 of image size
\item Rotate between -45 degrees and 45 degrees

\end{itemize}
The metric used to choose the best checkpoint was DICE score. The raw output of the model is a 1024x1024 matrix where each value corresponds to a sigmoid output between 0 and 1, associated with the probability of belonging to a lung area with finding (pneumothorax and pleural effusion in each module respectively). The sum of all values was thresholded to obtain a binary output.\\
For the pleural effusion model, epoch 20 of scheme 3 was chosen, and a threshold value of 3440.5 was used to binarize the sum of the raw output. For the pneumothorax model, epoch 12 of scheme 4 was chosen, and a threshold value of 144.43 was used to binarize the sum of the raw output. 
\subsection*{Inception-ResnetV2: Detection of lung opacities}
This architecture, published by Szegedy et al. [7], was implemented in Keras using Tensorflow as backend. Convolutional weights were transferred from a model pretrained on the the NIH ChestXray14 dataset, available at \cite{noauthor_undated-fl}. Three final layers were added: 
\begin{itemize}
    \item Fully connected layer of 256 units with ReLu activation. 
\item Dropout of p=0.5.
\item Fully connected layer of two units with softmax activation.

\end{itemize}

Input image size was 256x256, in batches of 32 images. We used Adam  as optimizer, with $\beta_1=0.9$ and $\beta_2=0.999$, and a learning rate of 1e-4. Categorical cross entropy was used as loss function and accuracy as validation metric. Data augmentation was applied in the same way as with AlbuNet-34.\\
Class activation map of opacity class, obtained using keras-visualization package. The raw output of the model is a pair of values that sum up to one, corresponding to the two softmax units. \\
The model chosen corresponds to epoch 17, using 2410 steps per epoch. To binarize the output, the value corresponding to opacity class was thresholded with a cutting point of 0.98.

\subsection*{RetinaNet: Detection of fractures}
This architecture, published by Lin et al. \cite{Lin2017-bf}, was implemented in Keras using Tensorflow as backend. Weights of Resnet50 backbone were transferred from a checkpoint of the lung opacity module. Input image size is determined so that the largest side is smaller than 1333 pixels, and the smaller side is larger than 800 pixels, keeping aspect ratio. This model is trained with focal loss, using Adam optimizer with a learning rate of 1e-5. Data augmentation applied included rotations, shear, scaling, and translation.  \\
The model chosen was trained for 13 epochs, using 4000 steps per epoch. The raw output of the model is a list of detected bounding boxes, each defined by five values: two pairs of coordinates and their corresponding confidence score. The maximum confidence score is thresholded to obtain one binary output per image, and we set a threshold value of 0.15.  

\section*{C. Detailed results}
\renewcommand{\thefigure}{C\arabic{figure}}
\renewcommand{\thetable}{C\arabic{table}}

Table \ref{tab:tableC1} presents the results of a basic demographic analysis of patients included in the local test set, organized by type of finding. Table \ref{tab:tableC2} shows the AUROC values observed for TRx detection for each  type of finding, evaluated in both test sets, as well as their bootstrap confidence intervals.

\newcolumntype{G}{>{\columncolor{gray!20!white}}p{0.2\textwidth}}
\begin{table}[h]
 \caption{Local test set demographic information by findings subsets.}
  \centering
 \begin{tabularx}{\textwidth}{GX|XX|XXX}
    \toprule
    
    \multirow{2}{*}{ } & \multirow{2}{*}{\shortstack{Number of \\images ($\%$) }} & \multicolumn{2}{c}{ Sex } & \multicolumn{3}{|c}{ Age } \\
& & \centering Female & \centering Male & Mean $\pm$ Std & Range & Median[IQR] \\ \midrule
All & 1064 & 543 (51$\%$) & 521(49$\%$) & 58 $\pm$  21 & 2-96 & 48 [32-64] \\ \midrule
Normal
(no finding) & 703 (66$\%$) & 368 (52$\%$) & 335 (48$\%$) & 44 $\pm$ 17 & 16-96 & 43 [31-58] \\ \hline
Abnormal & 361 (34$\%$) & 175 (48$\%$) & 186 (52$\%$) & 57 $\pm$ 22 & 2-95 & 61 [37-76] \\
\hspace{3mm} Pneumothorax & 119 & 42 & 77 & 41 $\pm$ 20 & 2-86 & 30 [24-57] \\
\hspace{3mm}Pleural effusion & 86 & 40 & 46 & 55 $\pm$ 21 & 2-91 & 58 [36-71] \\
\hspace{3mm}Lung opacity & 176 & 90 & 86 & 62 $\pm$ 20 & 10-94 & 66 [50-78] \\
\hspace{3mm}Fracture & 66 & 45 & 21 & 72 $\pm$ 14 & 36-95 & 75 [61-82] \\
 \bottomrule

\end{tabularx}
  \label{tab:tableC1}
\end{table}

\newcolumntype{g}{>{\columncolor{gray!20!white}}p{0.1\textwidth}}

\begin{table}[h]
 \caption{Area under the ROC curve for both test sets.}
  \centering
 \begin{tabularx}{\linewidth}{gXXXXX}
    \toprule
& Abnormality & Pneumothorax & Pleural Effusion & Lung opacity & Fracture \\ \midrule
\shortstack[l]{External \\test set} & 0.749 [0.74-0.76] & 0.888
[0.87-0.90] & - & 0.749[0.74-0.76] & 0.592 [0.56-0.62] \\ \hline
\shortstack[l]{Local \\test set} & 0.874 [0.86-0.89] & 0.901 [0.87-0.93] & 0.883 [0.85-0.91] & 0.793 [0.77-0.82] & 0.689 [0.64-0.74] \\
 \bottomrule

\multicolumn{6}{l}{\footnotesize{*Values in brackets correspond to 95$\%$ confidence intervals.
}}
\end{tabularx}
  \label{tab:tableC2}
\end{table}

\begin{figure*}[h]
  \includegraphics[width=\textwidth,keepaspectratio]{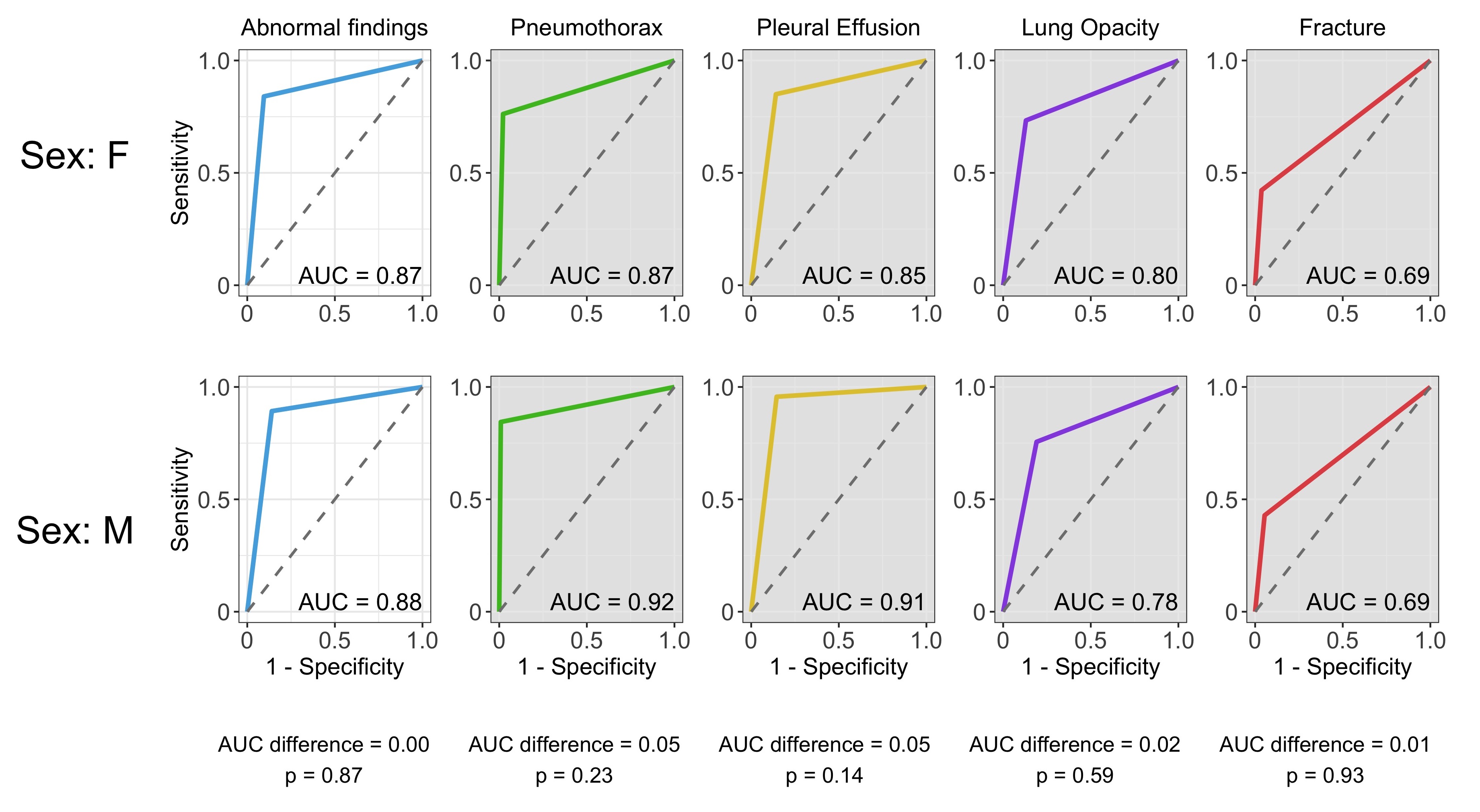}
  \centering
  %\fbox{\rule[-.5cm]{4cm}{4cm} \rule[-.5cm]{4cm}{0cm}}
  \caption{ROC curves by biological sex groups for the local test set. No significant differences were observed.}
  \label{fig:figC1}
\end{figure*}

\begin{figure*}[h]
  \includegraphics[width=\textwidth,keepaspectratio]{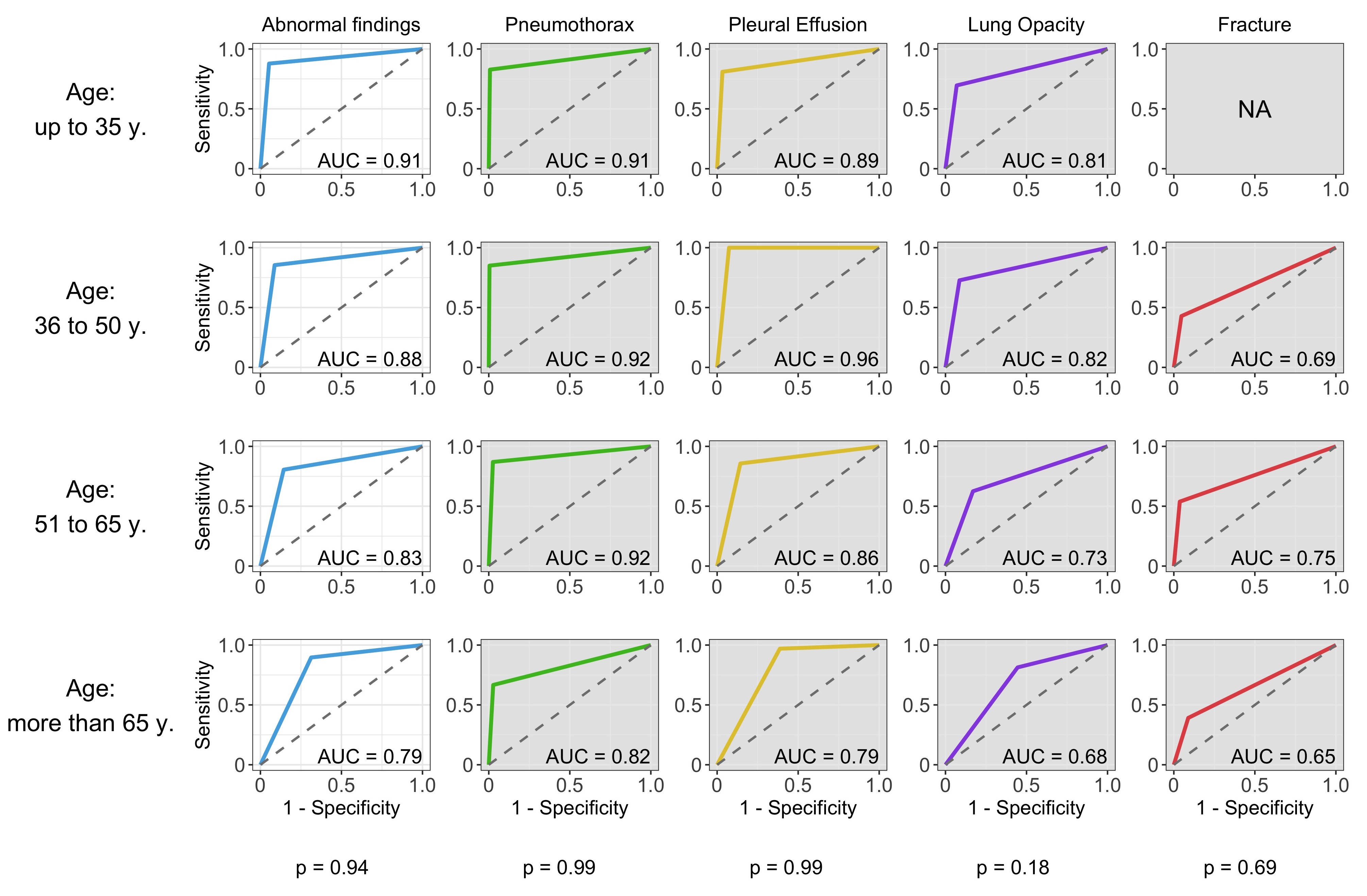}
  \centering
  %\fbox{\rule[-.5cm]{4cm}{4cm} \rule[-.5cm]{4cm}{0cm}}
  \caption{ROC curves by age groups for the local test set. No significant differences were observed. NA: Not available, due to insufficient images positive for fracture.}
  \label{fig:figC2}
\end{figure*}

\begin{figure*}[h]
  \includegraphics[width=\textwidth,keepaspectratio]{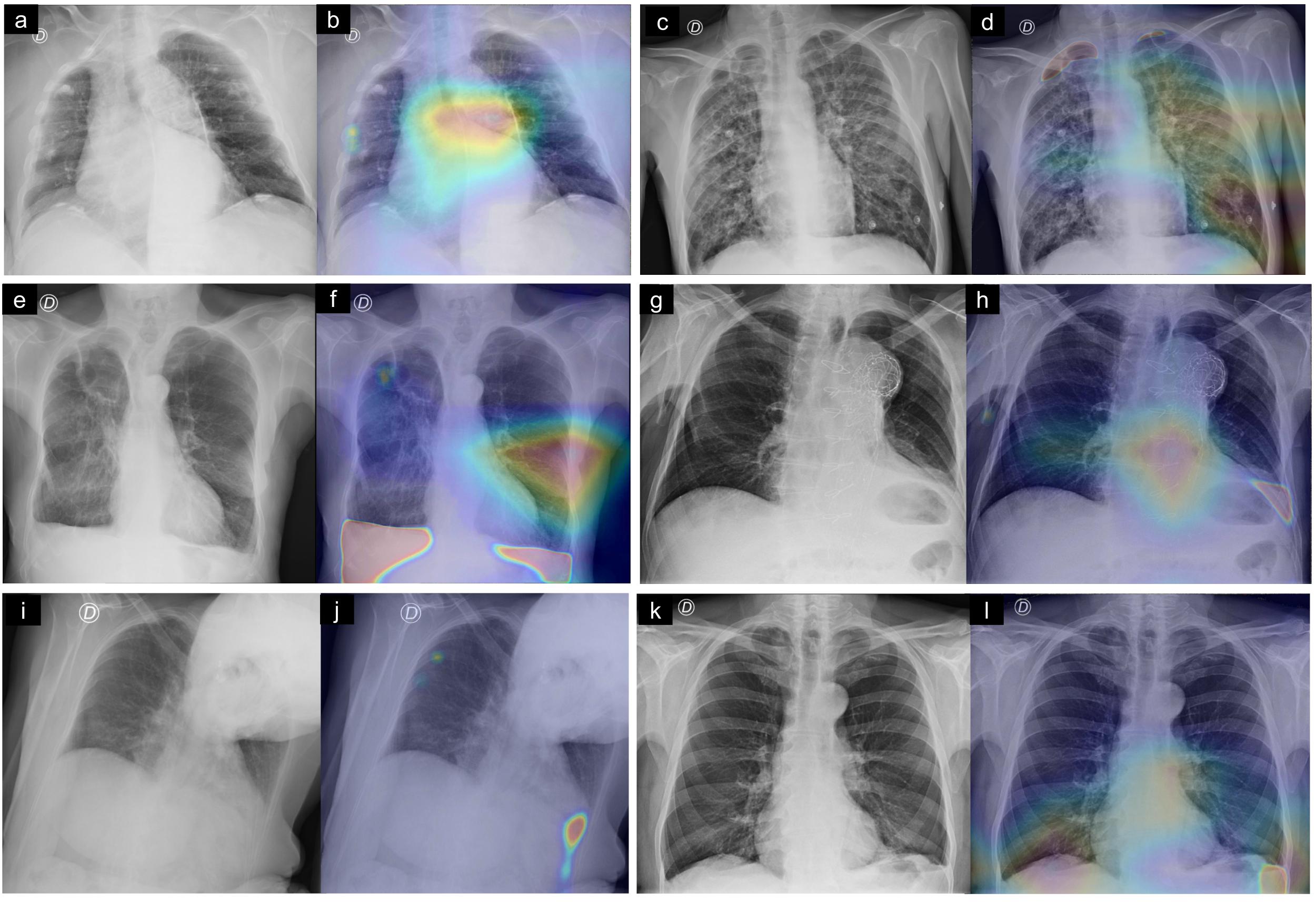}
  \centering
  %\fbox{\rule[-.5cm]{4cm}{4cm} \rule[-.5cm]{4cm}{0cm}}
  \caption{Examples of incorrect detections. Pairs of images are the original CXR (left) and its corresponding unified heatmap (right). (a,b) False lung opacity detected on mediastinum, and pleural calcifications  in the right middle rib cage are misinterpreted as fracture. (c,d) Subpleural apical bullae is confused with pneumothorax, and the activated opacities on hilar shadows could be prominent vessels. (e,f) Both lungs present multiple focal opacities, which are correctly detected in the left lung but misinterpreted as fractures in the upper right lung. Emphysema in both lower lungs causes blunting of the left costophrenic angle, which is incorrectly detected as pleural effusion. (g,h) Aortic prosthesis causes false positive detection of lung opacities and effusion. (i,j) Patient position is a technical difficulty for acquisition, with a rotated thorax that displaces cardiac silhouette to the left, causing opacities in the left costophrenic sinus that are confused with a pleural effusion. Rib fractures on the upper right rib cage are correctly detected. (k,j) Left pleural effusion is correctly detected, but there is a false positive detection for lung opacity on the lower right lung.}
  \label{fig:figC3}
\end{figure*}

For patients in the local test set, we performed a stratification by biological sex and by age group. TRx performance was evaluated for these subgroups. Figures \ref{fig:figC1} and \ref{fig:figC2} show ROC curves obtained. We calculated the difference between AUROCs among subgroups for each detection task, and performed a non-parametric test to compare the equality of ROC curves \cite{Martinez-Camblor2013-wt}. We observed no significant difference between groups, suggesting a fair performance.\\
We performed a qualitative revision of heatmaps to interpret TRx outputs and report its strengths and weaknesses. Figure \ref{fig:figC3} shows six cases where TRx had an incorrect finding detection, and the radiological analysis to understand the mistake in each case.
\bibliographystyle{unsrt}  
\bibliography{supplement}

